\documentclass[12pt,a4paper]{article}

\usepackage{amsmath,amsfonts,amssymb,bm}

\usepackage{graphicx} 

\usepackage[hang,normalsize,bf]{subfigure} 

\begin{document}

\begin{center}

{\Huge \bf From unintegrated gluon distributions 
to particle production in nucleon-nucleon collisions at RHIC energies} 
\footnote{To appear in the special issue of Acta Physica
  Polonica to celebrate the 65th Birthday of Prof. Jan Kwieci\'nski}

\vspace {0.6cm}

{\large A. Szczurek $^{1,2}$}

\vspace {0.2cm}

      $^{1}$ {\em Institute of Nuclear Physics\\
PL-31-342 Cracow, Poland\\}
$^{2}$ {\em University of Rzesz\'ow\\
PL-35-959 Rzesz\'ow, Poland\\}

\end{center}

\begin{abstract}
The inclusive distributions of gluons and pions are calculated
with absolute normalization for high-energy nucleon-nucleon
collisions. The results for
several unintegrated gluon distributions from the literature
are compared. The gluon distribution proposed recently
by Kharzeev and Levin based on the idea of gluon saturation
is tested against DIS data from HERA. 
We find huge differences in both rapidity and transverse
momentum distributions of gluons and pions in nucleon-nucleon
collisions for different models of unintegrated gluon distributions.
The approximations used recently in the literature
are discussed. The Karzeev-Levin gluon distribution gives
extremely good description of momentum distribution
of charged hadrons at midrapidities.
Contrary to a recent claim in the literature,
we find that the gluonic mechanism discussed does not describe
the inclusive spectra of charged particles in the fragmentation
region, i.e. in the region of large $|y|$ for any unintegrated
gluon distribution from the literature.
\end{abstract}

\section{Introduction}

The recent results from RHIC (see e.g. \cite{RHIC}) have attracted
renewed interest in better understanding the dynamics of
particle production, not only in nuclear collisions.

Quite different approaches \cite{thermal,PHOBOS,KL01} have been used
to describe the particle spectra from the nuclear collisions \cite{PHOBOS}.
The thermal models do not make a direct link to nucleon-nucleon
collisions. In contrast, in dual parton approaches (DPM)
the nucleon-nucleon collisions are the basic ingredients of
nuclear collisions.
Somewhat extreme model in Ref.\cite{KL01} with an educated guess
for unintegrated gluon distribution describes surprisingly well
the whole charged particle rapidity distribution by means
of gluonic mechanisms only. Such a gluonic mechanism would lead to
the identical production of positively and negatively charged hadrons.
The recent results of the BRAHMS experiment \cite{BRAHMS} put into
question the successful description of Ref.\cite{KL01} and
show that the DPM type approaches seems more correct.
In the light of the BRAHMS experiment is becomes obvious that
the large rapidity regions have more complicated flavour structure.
The pure gluonic mechanisms, if at all, can be dominant only at
midrapidities although the charged kaons \cite{BRAHMS} show that even
this is doubtful. Similarly also the thermal models have difficulties
to describe the (pseudo)rapidity dependence of particle to antiparticle
ratios \cite{BRAHMS} and have to limit to the midrapidity only.
In principle, the dynamics in nucleus-nucleus collision is
fairly complicated and requires a separate analysis.
In the following I concentrate only on nucleon-nucleon collisions --
the basic ingredients of the nucleus-nucleus collisions.

On the microscopic level the approach of Kharzeev and Levin
\cite{KL01} is based on the gluon-gluon fusion. The gluon-gluon
fusion is expected to be the dominant process at midrapidities and at
asymptotically large energies. It is not clear how large
the energy should be to validate this thesis. The physics
in the fragmentation region is somewhat different.
It was suggested long ago \cite{DH77} that pions
in the fragmentation region are correlated with the valence
quark distributions in hadrons.

The standard hadronization approaches are based rather on
the 2 $\rightarrow$ 2 partonic subprocesses which constitute
only a part of the dynamics. The perturbative
component of these hybrid models has a flavour structure as
dictated by the quark/antiquark distributions. On the other hand,
the flavour structure of the remaining soft component is
not so explicit.
Furthermore the partition into "soft" and "hard" components is
somewhat arbitrary, being to some extend rather an artifact
of a natural failure in applying the (2 $\rightarrow$ 2) pQCD
at low transverse momenta of hadrons than a clear border of
the two regions.

In this paper I discuss the relation between unintegrated gluon
distributions in hadrons and the inclusive momentum distribution
of particles produced in hadronic collisions.
The results obtained with different unintegrated gluon
distributions presented recently in the literature are shown
and compared.
In the present study I limit to the
nucleon-nucleon collisions only and leave the
nucleus-nucleus collisions for a separate analysis.

\section{Photon-nucleon cross section at high energies}

It became a standard in recent years to first describe the HERA data
and only then to test the resulting gluon distributions in other
processes. We try to follow this reasonable methodology also
for jet and particle production.

It is known that the LO total $\gamma^* N$ cross section
can be written in the form
\begin{equation}
\sigma_{tot}^{\gamma^* N} = \sum_q \int dz \int d^2 \rho
\; | \Psi_{\gamma^* \rightarrow q \bar q}(Q,z,\rho) |^2
\cdot \sigma_{(q \bar q) N}(x,\rho) \; .
\label{gamma_N}
\end{equation}
In this paper we take the so-called quark-antiquark
photon wave function of the perturbative form \cite{NZ90}.
As usual, in order to correct the photon wave function for large
dipole sizes (nonperturbative region) we introduce
an effective quark/antiquark mass ($m_{eff} = m_0$).
The dipole-nucleon cross section can be parametrized
or calculated from the unintegrated gluon distribution
\begin{eqnarray}
\sigma_{(q \bar q) N}(x,\rho)
&=&
\frac{4 \pi}{3} \int \frac{d^2 \kappa_t}{\kappa_t^2}
\left[1 - \exp(i\vec{\kappa_t} \vec{\rho}) \right]
\alpha_s {\cal F}(x,\kappa_t^2) \nonumber \\
&=& \frac{4 \pi^2}{3} \int \frac{d \kappa_t^2}{\kappa_t^2}
\left[ 1 - J_0(\kappa_t \rho) \right]
\alpha_s {\cal F}(x,\kappa_t^2) \; .
\label{Fourier_transform}
\end{eqnarray}
In the equation above the running coupling constant is fixed constant
or is frozen according to an analytic prescription \cite{SS97}.
In the next section we shall compare the dipole-nucleon cross sections
calculated from different unintegrated gluon distributions.

\section{Unintegrated gluon distributions}

Search for the unintegrated gluon distribution in the nucleon
was a subject of active both theoretical and phenomenological
research in recent years.
Still at present the unintegrated gluon distributions are rather
poorly known. The main reason of the difficulties is the fact
that the unintegrated gluon distribution is a quantity
which depends on at least two variables ($x$ and $\kappa^2$)
in a nontrivial and a priori unknown way.
Another difficulty is in an unambiguous separation of perturbative
and nonperturbative regions.
In general different phenomena test the unintegrated
gluon distribution in different corners of the phase space.
Therefore it is not surprising that different gluon distributions
found in the literature, extracted from the analyses of different
phenomena, differ among themselves considerably
\cite{small_x_collaboration}.
In this section I collect and briefly discuss gluon distributions
used in the present calculation of the jet and particle production.
There are two different conventions of introducing unintegrated
gluon distributions in the literature. The resulting quantities
are denoted as $f$ (dimenionless quantity)
and ${\cal F}$ (with dimension 1/GeV$^2$). We shall keep
this notation throughout the present paper.

\subsection{BFKL gluon distribution}

At very low $x$ the unintegrated gluon distributions are believed
to fulfil BFKL equation \cite{BFKL} (see also \cite{KMSR90}).
After some simplifications \cite{AKMS94} the BFKL equation reads
\begin{equation}
-x \frac{\partial f(x, q_t^2)}{\partial x} =
\frac{\alpha_s N_c}{\pi} q_t^2
\int_0^{\infty} \frac{dq_{1t}^2}{q_{1t}^2}
\left[                                 
\frac{f(x,q_{1t}^2) - f(x,q_t^2)}{|q_t^2 - q_{1t}^2|}
+ \frac{f(x,q_t^2)}{\sqrt{q_t^4+4 q_{1t}^4}}
\right] \; .
\label{BFKL_equation}
\end{equation}
The homogeneous BFKL equation can be solved numerically \cite{AKMS94}.
Here in the practical applications we shall use a simple
parametrization for the solution \cite{ELR96}
\begin{equation}
f(x,\kappa_t^2) = \frac{C}{x^{\lambda}}
\left(\frac{\kappa_t^2}{q_0^2}\right)^{1/2}
\frac{{\tilde \phi}_0}{\sqrt{ 2 \pi \lambda''\ln(1/x)}}
\exp\left[ - \frac{\ln^2(\kappa_t^2/{\bar q}^2)}{r2\lambda ''\ln(1/x)}
\right]
\label{ELR_parametrization}
\end{equation}
In the above expression $\lambda = 4 {\bar \alpha}_s \ln2$,
$\lambda''$ = 28 ${ \bar \alpha}_s \zeta(3)$, ${\bar \alpha}_s
= 3 \alpha_s/\pi, \zeta(3)$ = 1.202. The remaining parameters were
adjusted in \cite{ELR96} to reproduce with a satisfactory accuracy
the gluon distribution which was obtained in \cite{AKMS94}
as the numerical solution of the BFKL equation. 
It was found that ${\bar q} = q_0$ = 1, $C {\bar \phi_0}$ = 1.19
and $r$ = 0.15 \cite{ELR96}.

\subsection{Golec-Biernat-W\"usthoff gluon distribution}

Another parametrization of gluon distribution in the proton
can be obtained based on the Golec-Biernat-W\"usthoff
parametrization of the dipole-nucleon cross section with
parameters fitted to the HERA data \cite{GBW}.
The resulting gluon distribution reads \cite{GBW_glue}:
\begin{equation}
\alpha_s {\cal F}(x,\kappa_t^2) =
\frac{3 \sigma_0}{4 \pi^2} R_0^2(x) \kappa_t^2 \exp(-R_0^2(x)
\kappa_t^2) \; ,
\label{GBW_glue}
\end{equation}
where
\begin{equation}
R_0(x) = \frac{1}{GeV} \left( \frac{x}{x_0} \right)^{\lambda/2}
\; .
\end{equation}
From their fit to the data: $\sigma_0$ = 29.12 mb,
$x_0$ = 0.41 $\cdot$ 10$^{-4}$,
$\lambda$ = 0.277 \cite{GBW}.
In order to determine the gluon distribution needed in calculating
jet and particle production we shall take
$\alpha_s$ = 0.2.

\subsection{Kharzeev-Levin gluon distribution}

Another parametrization, also based on the idea of gluon
saturation, was proposed recently in \cite{KL01}.
In contrast to the GBW approach \cite{GBW}, where
the dipole-nucleon cross section is parametrized,
in the Karzeev-Levin approach it is the gluon distribution
which is parametrized.
In the following we shall consider the most simplified
functional form:
\begin{eqnarray}
{\cal F}(x,\kappa^2) =
\begin{cases}
 f_0                              & \text{if} \; \kappa^2 < Q_s^2 , \\
 f_0 \cdot \frac{Q_s^2}{\kappa^2} & \text{if} \; \kappa^2 > Q_s^2 .
\end{cases}
\label{KL_glue}
\end{eqnarray}
The saturation momentum $Q_s$ is parametrized exactly as in the GBW
model $Q_s^2(x) =$ 1 GeV$^2$ $\cdot \left( \frac{x_0}{x}
\right)^{\lambda}$. It was claimed in \cite{KL01} that the gluon
distribution like (\ref{KL_glue}) leads to a good description
of the recent RHIC rapidity distributions. It is interesting
to check its performance for the deep inelastic scattering
at low Bjorken $x$.

\begin{figure}[htbp] 
  \begin{center}
    \includegraphics[width=7cm]{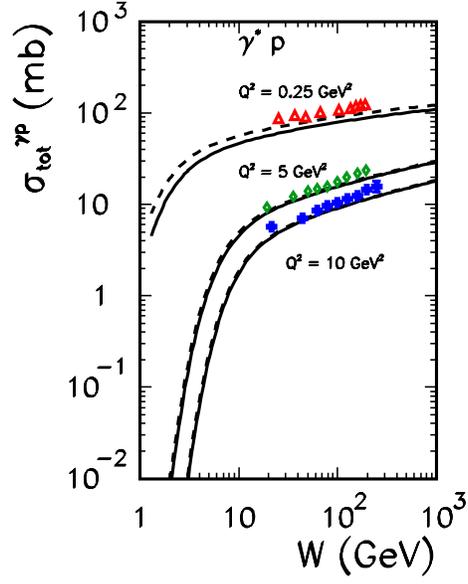}
    \caption{\it
      The cross section $\sigma_{tot}^{\gamma^* p}$ as a function of
      the center of mass energy $W$ for $Q^2$ = 0.25 GeV$^2$, $Q^2$ =
      5 GeV$^2$ and $Q^2$ = 10 GeV$^2$.  The results obtained with the
      KL gluon distribution ($m_0$ = 0.15/0.10 GeV) are shown by the
      solid and dashed lines.  Experimental data were taken from
      \cite{HERA_data}.
      \label{fig:HERA}}
  \end{center}
\end{figure}

In the following the normalization constant $f_0$ is adjusted
to roughly describe the HERA data. We find $f_0$ = 170 mb.
The quality of the fit is shown in Fig.~\ref{fig:HERA} for
$Q^2$ = 0.25, 5, 10 GeV$^2$.
In this fit the running coupling constant frozen according to
\cite{SS97} was used. The result at low photon virtuality
($Q^2$ = 0.25 GeV$^2$) depends also on the value of the quark/antiquark
effective mass. In the calculation in Fig.~\ref{fig:HERA}
$m_0$ = 0.15 GeV (solid) and $m_0$ = 0.10 GeV (dashed) was used.
It can be inferred from the figure that the
(virtual) photon-proton cross section at large virtuality
($Q^2$ = 5,10 GeV$^2$) is in practice independent of the effective
quark mass. This allows to fix the gluon normalization
constant~$f_0$.

\begin{figure}[htbp] 
  \begin{center}
    \includegraphics[width=7cm]{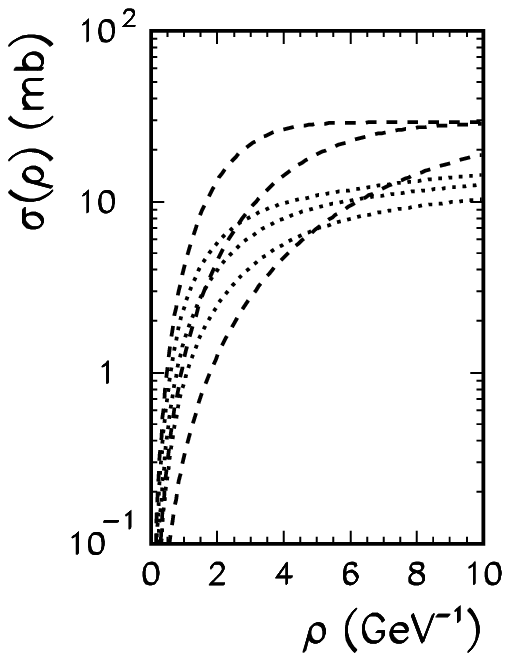}
    \caption{\it
      The dipole-nucleon cross section as a function of the transverse
      dipole size
      $\rho$ for the GBW (dashed) and KL (dotted)
      unintegrated gluon distributions.
      \label{fig:dip_nuc}}
  \end{center}
\end{figure}

In order to better visualize the difference to the GBW model,
in Fig.~\ref{fig:dip_nuc} I compare the dipole-nucleon
cross sections in both parametrizations
for $x=10^{-2}$, 10$^{-3}$, 10$^{-4}$.
In the GBW approach the dipole-nucleon cross section saturates
at large dipole size. In contrast to the GBW
parametrization, the dipole-nucleon cross section calculated
according to Eq.(\ref{Fourier_transform}) based on
the KL gluon distribution (\ref{KL_glue}) grows slowly with
the dipole size $\rho$.

\subsection{Kimber-Martin-Ryskin gluon distribution}

The unintegrated gluon distribution can be obtained even
when the integrated gluon distribution fulfils standard
DGLAP evolution equation.
At very small~$x$
\begin{equation}
{\cal F}(x,\kappa^2) = \frac{\partial}{\partial Q^2}
\left[ x g(x,Q^2) \right] |_{Q^2 = \kappa^2} \; .
\label{derivative}
\end{equation}
This prescription breaks at larger values of $x$ when
the derivative of the gluon distribution becomes negative.
This may be somewhat improved by introducing a Sudakov form factor
$T_g(\kappa^2,\mu^2)$.
Then the unintegrated gluon distribution reads \cite{KMR}:
\begin{equation}
{\cal F}(x,\kappa^2,\mu^2) =
\frac{\partial}{\partial Q^2}
\left[
T(Q^2,\mu^2) x g(x,Q^2) \right] |_{Q^2 = \kappa^2}
 \; .
\label{KMR}
\end{equation}

Resumming virtual contributions to DGLAP equation,
the unintegrated parton distributions can be written as \cite{KMR}
\begin{equation}
f_a(x,\kappa^2,\mu^2) = T_a(\kappa^2,\mu^2)
\cdot \frac{\alpha_s(\kappa^2)}{2 \pi}
\sum_{a'} \int_x^{1-\delta} P_{aa'}(z) \left(\frac{x}{z} \right)
a'\left(\frac{x}{z},\kappa^2\right) dz \; .
\label{KMR_master} 
\end{equation}
Specializing to the gluon distribution the Sudakov form factor
reads as
\begin{equation}
T_g(\kappa^2,\mu^2) =
\exp\left(       
-\int_{\kappa^2}^{\mu^2} \frac{d p^2}{p^2} \frac{\alpha_s(p^2)}{2 \pi}
\int_{0}^{1-\delta} dz z \left[P_{gg}(z)+\sum_q P_{qg}(z)     \right]
\right) \; .
\label{Sudakov}
\end{equation}
The Sudakov form factor introduces a dependence on a second scale $\mu^2$.
It is reasonable to assume that the unintegrated gluon density
given by (\ref{KMR_master}) starts only
for $\kappa_t^2 > \kappa_{t0}^2$ \cite{KMS97}.
At lower $\kappa_t^2$ an extrapolation is needed.
In our case of particle distributions the results
are sensitive to rather low $\kappa$. Because of this, a use of
the GRV integrated gluon distribution \cite{GRV95,GRV98}
in (\ref{KMR_master}) seems more adequate than any other PDF.
Following Ref.\cite{MT02} $\kappa_{t0}^2$ = 0.5 GeV$^2$ is taken
as the lowest value where the unintegrated gluon distribution is
calculated from Eq.(\ref{KMR_master}). Below it is assumed
\begin{equation}
{\cal F}(x,\kappa^2) = f(x,\kappa^2)/\kappa^2 =
f(x,\kappa_0^2)/\kappa_0^2 \; .
\label{low_kappa}
\end{equation}

The choice of $\mu^2$ in our case of jet (particle) production
is not completely obvious.
In the present analysis $\mu^2 = p_t^2$ is assumed,
where $p_t$ is transverse momentum of the produced gluon ($\equiv$ jet).
In accord with the interpretation of the Sudakov form factor as
a survival probability we assume that if transverse momentum of
the produced gluon is smaller than
the transverse momentum of the last gluon of the ladder
($p_t < \kappa_{1}$ or $p_t < \kappa_{2}$, see next section)
then the corresponding Sudakov form factor is set to 1, i.e. we do not
allow for any enhancement.
If $T_g$ in Eq.(\ref{KMR_master}) is ignored we shall denote
the corresponding gluon distribution as $f_{DGLAP}$ or
${\cal F}_{DGLAP}$ and call it DGLAP gluon distribution for brevity.


\subsection{Bl\"umlein gluon distribution}

In the approach of Bl\"umlein \cite{Blue95}
the $\kappa_t^2$ dependent gluon distribution satisfying the BFKL
equation can be represented as the convolution of the integrated gluon
density $x g(x,\mu^2)$ and a universal function $\cal{B}$
\begin{equation}
{\cal F}(x,\kappa_t^2,\mu^2) = \int_x^1 {\cal B}(z,\kappa_t^2,\mu^2)
\frac{x}{z} g(\frac{x}{z},\mu^2) dz   \; .
\label{bluemlein_convo}
\end{equation}
The universal function ${\cal B}(x,\kappa_t^2,\mu^2)$ can be
represented as a series \cite{Blue95}.
The first term of the expansion describes BFKL dynamics
in the double-logarithmic approximation:
\begin{equation}
{\cal B}(z,\kappa_t^2,\mu^2) =
\begin{cases}
\frac{{\bar \alpha}_s}{z \kappa_t^2}
  J_0(2 \sqrt{{\bar \alpha_s} \log(1/z) \log(\mu^2/\kappa_t^2)})
                             & \text{if} \; \kappa_t^2 < \mu^2 \\
\frac{{\bar \alpha}_s}{z \kappa_t^2}
  I_0(2 \sqrt{{\bar \alpha_s} \log(1/z) \log(\kappa_t^2/\mu^2)})
                             & \text{if} \; \kappa_t^2 > \mu^2 ,
\end{cases}
\label{B_function}
\end{equation}
where ${\bar \alpha_s} = 3 \alpha_s/\pi$.

In DIS there is a natural choice of the scale $\mu^2$.
The choice of the scale is not so obvious in the case
considered in the present paper. I shall argue that in practice
the dependence on that scale is very weak.
In the following I shall use the integrated gluon distribution
in Eq.(\ref{bluemlein_convo}) from Ref. \cite{GRV95}.

\section{Inclusive gluon production}

Before we go to particle production in the next section, let us
consider the first step of the process -- production of partons.
At high energies gluons are the most abundantly produced partons
in hadron-hadron collisions. Also gluons are responsible for
their production.

At sufficiently high energy the cross section for inclusive
gluon production in $h_1 + h_2 \rightarrow g$ can be written
in terms of the unintegrated gluon distributions ``in'' both colliding
hadrons:
\begin{equation}
\frac{d \sigma}{dy d^2 p_t} = \frac{16  N_c}{N_c^2 - 1}
\frac{1}{p_t^2}
\int
 \alpha_s(\Omega^2)
 {\cal F}_1(x_1,\kappa_1^2) {\cal F}_2(x_2,\kappa_2^2)
\delta(\vec{\kappa}_1+\vec{\kappa}_2 - \vec{p}_t)
\; d^2 \kappa_1 d^2 \kappa_2    \; .
\label{inclusive_glue0}
\end{equation}
In the equation above $f_1$ and $f_2$ are unintegrated gluon
distributions in hadron $h_1$ and $h_2$, respectively.
The longitudinal momentum fractions are fixed by kinematics:
$x_{1/2} = \frac{p_t}{\sqrt{s}} \cdot \exp(\pm y)$.
Generally the smaller jet (parton) momenta $p_t$,
the smaller $x_{1/2}$ come into play.
The argument of the running coupling constant is taken as
$\Omega^2 = \max(\kappa_1^2,\kappa_2^2,p_t^2)$.
The formula (\ref{inclusive_glue0}) above was
first written by Gribov, Levin and Ryskin
\cite{GLR81} (see also \cite{LL94}) and used later e.g. in \cite{ELR96}.
As discussed in Ref.\cite{GM97} the normalization of the cross section
in some previous works was not always correct.
Making use of the $\delta$ function (momentum conservation) one can
simplify (\ref{inclusive_glue0}) to the integral
\begin{equation}
\frac{d \sigma}{dy d^2 p_t} = \frac{16 N_c}{N_c^2 - 1}
\frac{1}{p_t^2} {\frac{1}{4}}
\int
 \alpha_s(\Omega^2)
{\cal F}_1\left(x_1,\left( \frac{\vec{p}_t + \vec{q}_t}{2} \right) \right)
{\cal F}_2\left(x_2,\left( \frac{\vec{p}_t - \vec{q}_t}{2} \right) \right)
 d^2 q_t 
 \; ,
\label{inclusive_glue1}
\end{equation}
where $\vec{q}_t = \vec{\kappa}_1 - \vec{\kappa}_2$ was introduced.
The factor 1/4 is the jacobian of transformation from
($\vec{\kappa}_1$, $\vec{\kappa}_2$) to ($\vec{p}_t$, $\vec{q}_t$).
The integral above is a two-dimensional integral over $d^2 q_t$,
i.e. over $q_t dq_t d\phi$,
where $\phi$ is the azimuthal angle between $q_t$ and $p_t$.
The original integral (\ref{inclusive_glue1}) can be written as
\begin{equation}
\frac{d \sigma}{dy d^2 p_t} = \int \; I(\phi) \; d \phi \; ,
\label{I_phi_def}
\end{equation}
where
\begin{equation}
I(\phi) =
\frac{4 N_c}{N_c^2 - 1}
\frac{1}{p_t^2}
\int
\alpha_s(\Omega^2)
{\cal F}_1\left(x_1,\kappa_1^2\right)
{\cal F}_2\left(x_2,\kappa_2^2\right)
 \; q_t d q_t     \; .
\label{I_phi_explicit}
\end{equation}
%

\begin{figure}[htb] 
  \begin{center}
    \includegraphics[width=8cm]{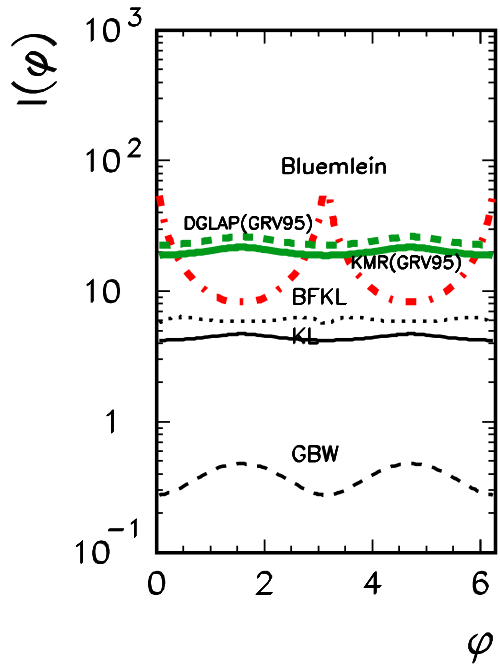}
    \caption{\it
      The intrinsic azimuthal correlations for different unintegrated
      gluon distributions: GBW (dashed), KL (solid), BFKL (dotted),
     Bl\"umlein (thick dash-dotted),
     DGLAP (thick dashed) and
      KMR (thick solid) at W = 200 GeV.  
      \label{fig:phi_correl}}
  \end{center} 
\end{figure}

In Fig.~\ref{fig:phi_correl}, I show the intrinsic angular correlation
function~$I(\phi)$
for different models of unintegrated gluon distributions for a RHIC
energy $W$ = 200 GeV. In this calculation y=0 and
$p_t$ = 1 GeV was taken. Quite a different pattern is obtained
for different unintegrated gluon distributions.
The $\phi$-distribution is flat for the KL, BFKL, DGLAP and KMR
gluon distributions.
The most pronounced structure is obtained with the Bl\"umlein gluon
distribution \cite{Blue95}(GRV95, $\mu^2$ = 10 GeV$^2$).
It was checked that the Bl\"umlein (GRV95) gluon distribution is not
very sensitive to the choice of the second scale $\mu^2$. 
The $\phi$ dependence at $y \ne$ 0 also strongly depends
on the unintegrated gluon distribution.

It was suggested in \cite{KL01} that the integral
(\ref{inclusive_glue1}) may be approximated by the formula
\begin{equation}
\frac{d \sigma}{dy d^2 p_t} = \frac{4 N_c \alpha_s}{N_c^2 - 1}
\frac{1}{p_t^2}
\int \left[ {\cal F}_1(x_1,{p}_t^2) {\cal F}_2(x_2,{q}_t^2) +
            {\cal F}_1(x_1,{q}_t^2) {\cal F}_2(x_2,{p}_t^2) \right]
 d q_t^2    \; .
\label{inclusive_glue2}
\end{equation}
%

\begin{figure}[htb] 
  \begin{center}
    \includegraphics[width=8cm]{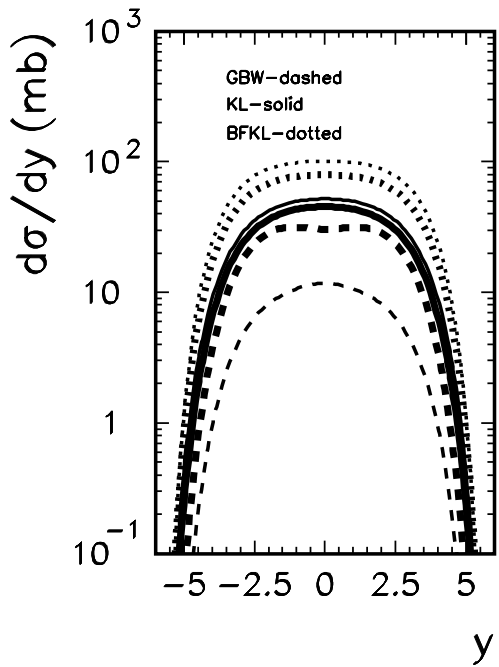}
    \caption{\it
      A comparison of the gluon rapidity distributions obtained from
      the exact (\ref{inclusive_glue1}) (thick lines) and approximate
      (\ref{inclusive_glue2}) (thin lines) formula for different models of
      unintegrated gluon distributions at W = 200 GeV.
      \label{fig:approx1}}
  \end{center}
\end{figure}

\begin{figure}[htb] 
\begin{center}
\includegraphics[width=8cm]{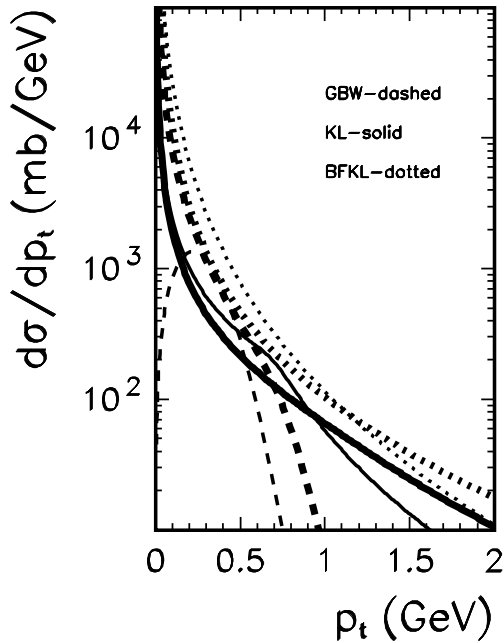}
\caption{\it
A comparison of the gluon transverse momentum distributions obtained
from the exact (\ref{inclusive_glue1}) (thick lines) and approximate
(\ref{inclusive_glue2})
(thin lines) formula for different models of unintegrated gluon
distributions at W = 200 GeV.
\label{fig:approx2}}
\end{center}
\end{figure}

In Fig.\ref{fig:approx1} (rapidity distribution) and
Fig.\ref{fig:approx2} (transverse momentum distribution)
I compare the results using the exact Eq.(\ref{inclusive_glue1})
and the approximate Eq.(\ref{inclusive_glue2}) formulae
for different models of unintegrated gluon distributions.
In Fig.\ref{fig:approx1} the integration over $p_t >$ 0.5 GeV
is performed while in Fig.\ref{fig:approx2} -1 $ < y < $ 1.
In both cases $\alpha_s$ was fixed at 0.2.
As can be seen by inspection of the figures
the use of the approximate formula is quantitatively justified
for the KL, BFKL gluon distributions and not justified
for the GBW one.


\begin{figure}[htbp] 
\begin{center}
\includegraphics[width=8cm]{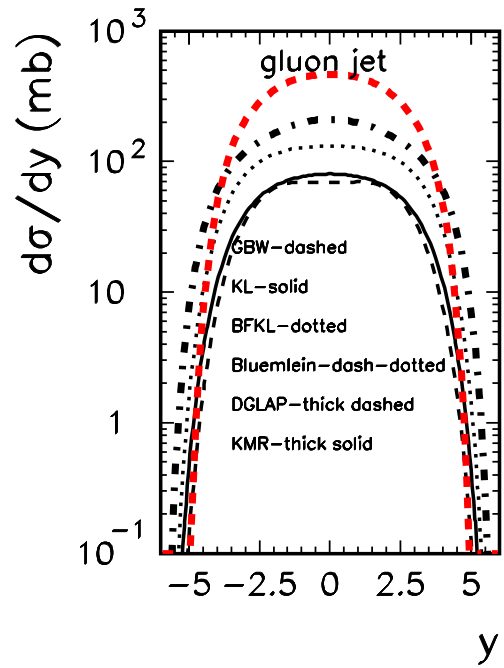}
\caption{\it
Inclusive gluon rapidity distribution
($p_t >$ 0.5 GeV) at W = 200 GeV 
for different models
of unintegrated gluon distributions.
\label{fig:glu_y}
}
\end{center}
\end{figure}

\begin{figure}[htbp] 
\begin{center}
\includegraphics[width=8cm]{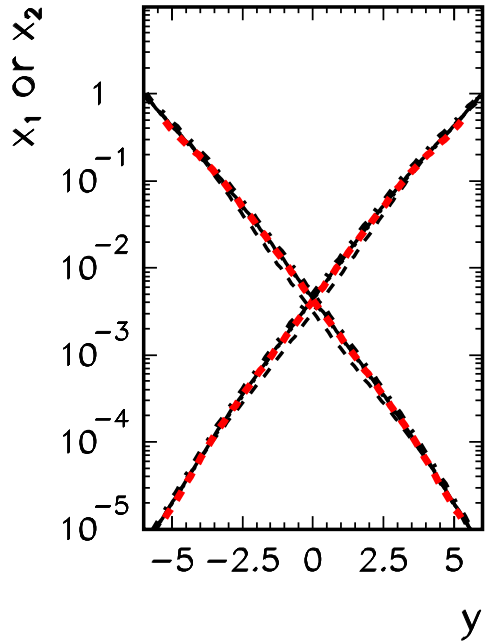}
\caption{\it
The average value of $x_1$ and $x_2$
for $p_t >$ 0.5 GeV and at W = 200 GeV. 
Lines corresponding to different unintegrated gluon PDF
are identical as in the previous figure.
\label{fig:x1_x2}
}
\end{center}
\end{figure}

In Fig.\ref{fig:glu_y} I compare the cross section
$\frac{d \sigma}{dy}(y)$ for different models of
unintegrated gluon distributions. In this calculation
$p_t <$ 0.5 was assumed. The rapidity distribution
of gluons are rather different for different gluon PDF.
Average values of $x_1$ and $x_2$ obtained with
different gluon distribution with the $p_t$ interval
chosen are shown in Fig.\ref{fig:x1_x2}. The following
general observations can be made. Average value $<x_1>$ and $<x_2>$
only weakly depend on the model of unintegrated gluon distribution.
For $y \sim$ 0 at the RHIC energy W = 200 GeV one tests
unintegrated gluon distributions at $x_g$ = 10$^{-3}$ - 10$^{-2}$.
This is the region known already from the HERA kinematics. 
When $|y|$ grows one tests more and more asymmetric (in $x_1$ and $x_2$)
configurations. For large $|y|$ either $x_1$ is extremely
small ($x_1 <$ 10$^{-4}$) and $x_2 \rightarrow$ 1
or $x_1 \rightarrow$ 1 and $x_2$ is extremely small ($x_2 <$ 10$^{-4}$).
These are regions of gluon momentum fraction where the unintegrated
gluon PDF is rather poorly known. The approximation used in obtaining
unintegrated gluon distributions are valid certainly only for $x <$ 0.1.
In order to extrapolate the gluon distribution to
$x_g \rightarrow$ 1 I multiply
the gluon distributions from the previous section by a factor
$(1-x_g)^n$, where n = 5-7.


In the approach considered in the present paper
(for details see next section)
the production of particles is sensitive to rather small gluon (called
equivalently jet despite of the small transverse momentum)
transverse momenta.

\begin{figure}[htb] 
\begin{center}
\includegraphics[width=8cm]{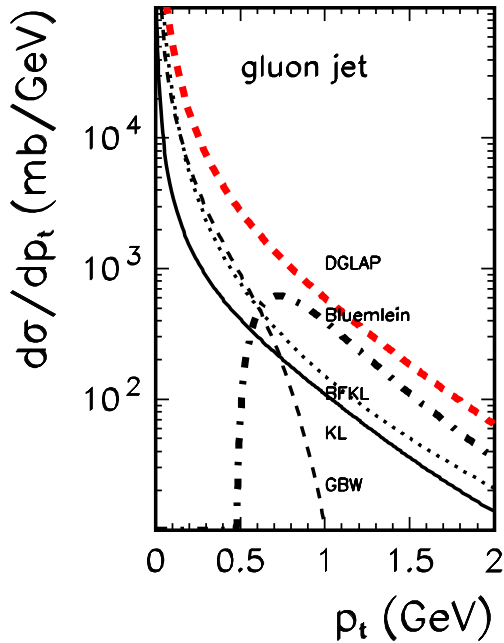}
\caption{\it
Inclusive gluon transverse momentum distribution 
(-1 $< y <$ 1) at W = 200 GeV for
different models of unintegrated gluon distributions:
BFKL (dotted), GBW (dashed), KL (solid), Bl\"umlein (thick
dash-dotted) and DGLAP (thick dashed).
\label{fig:glu_pt}
}
\end{center}
\end{figure}

In Fig.\ref{fig:glu_pt} I plot
$\frac{d \sigma}{dp_t}(p_t)$ in the low $p_t$ region.
In these calculations the gluon rapidity was integrated
in the interval -1 $< y <$ 1.
The results obtained with different models for unintegrated
gluon distributions differ considerably.
The transverse momentum distribution obtained
with the GBW gluon density is much steeper than
the distribution for any other gluon density.
The inclusion of DGLAP evolution as in \cite{BGK02}
would probably change the situation.
In the case of the Bl\"umlein gluon distribution
the transverse momentum spectrum has a natural low-$p_t$ cut-off
if the scale $\mu^2 = p_t^2$ is chosen. 
If similar prescription of the scale is used for calculating
gluon transverse momentum distribution with KMR method
the DGLAP and KMR results are almost identical.
Contrary to the claim in \cite{KL01} the result obtained
with the GBW and KL gluon distributions differ considerably.

The rapidity and pseudorapidity distributions of partons
(massless particles) are identical. The situation changes when
massive particles are produced in the final state via fragmentation.
Below we discuss how to take into account the unknown hadronization
process with the help of phenomenological fragmentation functions.

\section{From gluon to particle distributions}

In Ref.\cite{KL01} it was assumed, based on the concept
of local parton-hadron duality, that the rapidity distribution
of particles is identical to the rapidity distribution of gluons.
This seems to be a very severe assumption and for massive particles
this idea must lead to incorrect results, especially in the fragmentation
region. This approach leads to e.g. (massive) particles with rapidities
($y_h$) beyond the allowed kinematical region ($y_{h,min},y_{h,max}$).
Furthermore in \cite{KL01} the normalization of rapidity distributions
was fitted to the experimental charged particle rapidity distributions.
In our opinion, the good description of the charged particle
distribution in the full range of rapidity in Ref. \cite{KL01} is
due to these simplifications rather than due to
the underlying dynamics.

In the present approach I follow a different, yet simple, approach
which makes use of phenomenological fragmentation functions
(see e.g.\cite{W00,EH02}).
For our present exploratory study it seems
sufficient to assume that the emitted hadron, mostly pion,
is collinear to the gluon direction ($\theta_h = \theta_g$).
This is equivalent to $\eta_h = \eta_g = y_g$, where $\eta_h$ and
$\eta_g$ are hadron and gluon pseudorapitity, respectively.

In experiments a good identification of particles is not always
achieved which makes impossible to determine the rapidity of a particle.
The practice then is to measure pseudorapidity.
The rapidity of a given type of hadrons ($y_h$) with a mass $m_h$ can
be obtained from the pseudorapidity as
\begin{equation}
y_h = \frac{1}{2}
 \left[
\frac{\sqrt{\frac{m_h^2+p_{t,h}^2}{p_{t,h}^2} + \sinh^2\eta_h } + \sinh\eta_h }
     {\sqrt{\frac{m_h^2+p_{t,h}^2}{p_{t,h}^2} + \sinh^2\eta_h } - \sinh\eta_h }
 \right] \; .
\label{yh_etay}
\end{equation}
The collinearity of partons and particles leads to the following
relation between rapidity of the gluon and hadron
\begin{equation}
y_g = \mathrm{arsinh} \left( \frac{m_{t,h}}{p_{t,h}} \sinh y_h \right)
\; ,
\label{yg_yh}
\end{equation}
where the transverse mass $m_{t,h} = \sqrt{m_h^2 + p_{t,h}^2}$.
In order to introduce phenomenological fragmentation functions
one has to define a new kinematical variable.
In accord with $e^+e^-$ and $e p$ collisions I define a standard
auxiliary quantity $z$ by the equation $E_h = z E_g$.
This leads to the following relation between transverse momenta
of the gluon and hadron
\begin{equation}
p_{t,g} = \frac{p_{t,h}}{z} J(m_{t,h},y_h) \; ,
\label{ptg_pth}
\end{equation}
where
\begin{equation}
J(m_{t,h},y_h) =
\left( 1 - \frac{m_h^2}{m_{t,h}^2 \cosh^2 y_h} \right)^{-1/2} \; .
\label{J}
\end{equation}
Now we can write the single particle distribution
in terms of the gluon distribution from
the last section as follows
\begin{eqnarray}
\frac{d \sigma (\eta_h, p_{t,h})}{d \eta_h d^2 p_{t,h}} =
\int d y_g d^2 p_{t,g} \int 
dz \; D_{g \rightarrow h}(z,\mu_D^2) \\ \nonumber
\delta(y_g - \eta_h) \; 
\delta^2\left(\vec{p}_{t,h} - \frac{z \vec{p}_{t,g}}{J}\right)
\cdot \frac{d \sigma (y_g, p_{t,g})}{d y_g d^2 p_{t,g}} \; .
\label{from_gluons_to_particles}
\end{eqnarray}
Making use of the $\delta$ functions we can write
the single particle spectrum as
\begin{equation}
\frac{d \sigma(\eta_h,p_{t,h})}{d \eta_h d^2 p_{t,h}} =
\int_{z_{min}}^{z_{max}} dz
\frac{J^2 D_{g \rightarrow h}(z, \mu_D^2)}{z^2}
\frac{d \sigma(y_g,p_{t,g})}{d y_g d^2 p_{t,g}}
 \Bigg\vert_{y_g = \eta_h \atop p_{t,g} = J p_{t,h}/z} \; .
\label{single_particle_spectrum}
\end{equation}
Experimentally instead of the two-dimensional spectrum
(\ref{single_particle_spectrum}) one determines
rather one-dimensional spectra in either $\eta_h$ or $p_{t,h}$.

The one-dimensional pseudorapidity distribution can be obtained
by integration over hadron transverse momenta
\begin{equation}
\frac{d \sigma(\eta_h)}{d \eta_h} =
\int d^2 p_{t,h} \;
\frac{d \sigma(\eta_h,p_{t,h})}{d \eta_h d^2 p_{t,h}} \; .
\label{eta_had_distribution}
\end{equation}
%

\begin{figure}[htb] 

\begin{center}
\includegraphics[width=8cm]{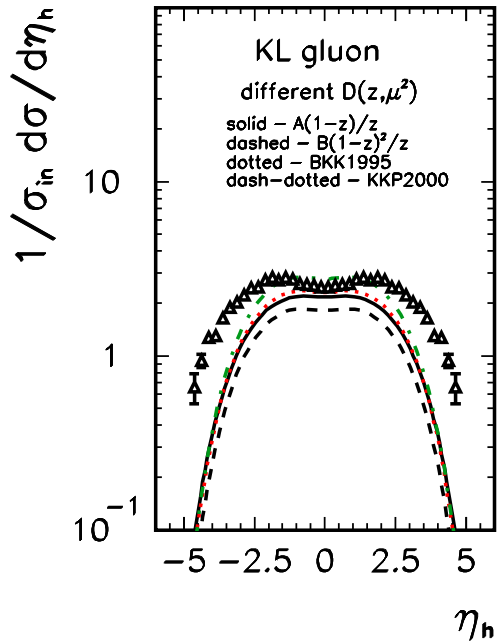}
\caption{\it
Charged-pion pseudorapidity distribution at W = 200 GeV
for the KL unintegrated gluon distribution
for different parametrizations of fragmentation functions.
In this calculation $p_{t,h} >$ 0.2 GeV.
The experimental data of the UA5 collaboration are taken
from \cite{UA5_exp}.
\label{fig:eta_ff}
}
\end{center}
\end{figure}

Stable particles \footnote{Here by stable particles we mean the particles
registered in detectors} are produced directly in the fragmentation
process or are decay products of other unstable particles.
There are a few global analyses of fragmentation function in
the literature up to next-to-leading order
\cite{BKK95,KKP00,Kretzer00,BFGW01}.
In the present calculation I shall use only leading order
fragmentation functions from \cite{BKK95,KKP00}.
One should remember, however, that both $e^+ e^-$ and $e p$ collisions
do not allow to uniquely determine $D_{g \rightarrow h}$ fragmentation
functions.
In order to test sensitivity of our results to these, in my opinion,
not quite well known objects I shall use also simple functional forms:
$D_{g\rightarrow h}(z) = 2 \frac{1-z}{z}$ (model I) or
$D_{g\rightarrow h}(z) = 3 \frac{(1-z)^2}{z}$ (model II) with
the factors in front adjusted to conserve momentum sum rule.
When charged particles are measured only, then to a good approximation
it is sufficient to multiply the fragmentation functions above
by a factor 2/3.

\begin{figure}[htb] 

\begin{center}
\includegraphics[width=8cm]{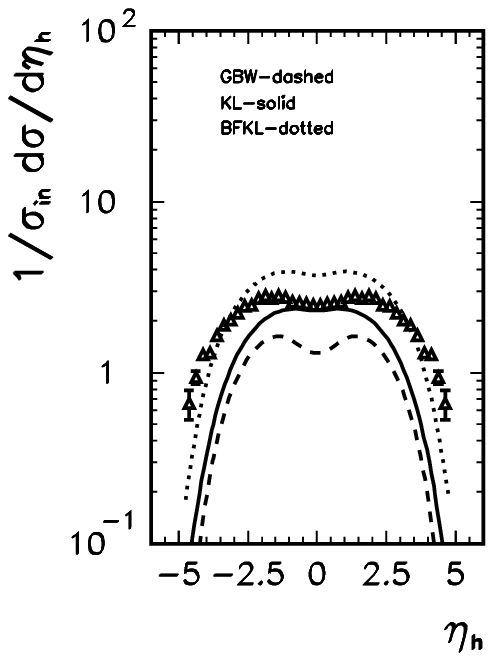}
\caption{\it
Charged-pion pseudrapidity distribution at W = 200 GeV
for different models of unintegrated gluon distributions.
In this calculation $p_{t,h} >$ 0.2 GeV.
The experimental data of the UA5 collaboration are taken
from \cite{UA5_exp}.
\label{fig:eta_glue}
}
\end{center}
\end{figure}

In Fig.\ref{fig:eta_ff} I compare pseudorapidity
distribution of charged pions at W = 200 GeV calculated with
the KL gluon distribution and different parametrizations of
fragmentation functions.
For the BKK1995 \cite{BKK95} and for the KKP2000 \cite{KKP00}
fragmentation functions the factorization scale was set to
$\mu_D^2 = p_{t,g}^2$, except for $p_{t,g} <$ 1 GeV, where
it was frozen at $\mu_D^2 =$ 1 GeV$^2$.
For reference shown are also experimental data for charged particles
measured by the UA5 collaboration at CERN \cite{UA5_exp}.
The results only weakly depend on the choice of
the $g \rightarrow \pi$ fragmentation function.
It is worth stressing that the theoretical cross section at
$\eta_h \approx$ 0 is almost consistent with the experimental one. 
However, the shapes of theoretical and experimental pseudorapidity
distributions differ significantly. It seems there is a room
for different mechanisms typical for fragmentation regions.
The specificity of these regions will be discussed elsewhere.

Let us analyze now how the results for pseudorapidity distributions
depend on the choice of the unintegrated gluon distribution.
In Fig.\ref{fig:eta_glue} I compare pseudorapidity distribution
of charged pions for different models of unintegrated
gluon distributions. In this calculation the Binnewies-Kniehl-Kramer
fragmentation function \cite{BKK95} has been used.
The conclusions inferred above stay true also here.
Having in view a dramatically steep $p_{t,g}$ distribution
in Fig.\ref{fig:glu_pt} it is rather surprising that the normalization
of the spectra at midrapidities comes roughly correct, although
very is a tendency
to an overestimation for some gluon distributions. This can be due to
the fit to DIS data,
where the resolved photon component has been neglected.
If the resolved photon component is explicitly included \cite{PS03}
then the normalization of the dipole-nucleon component
(dipole-nucleon cross section or unintegrated gluon distribution)
must be reduced.

\begin{figure}[htb] 

\begin{center}
\includegraphics[width=8cm]{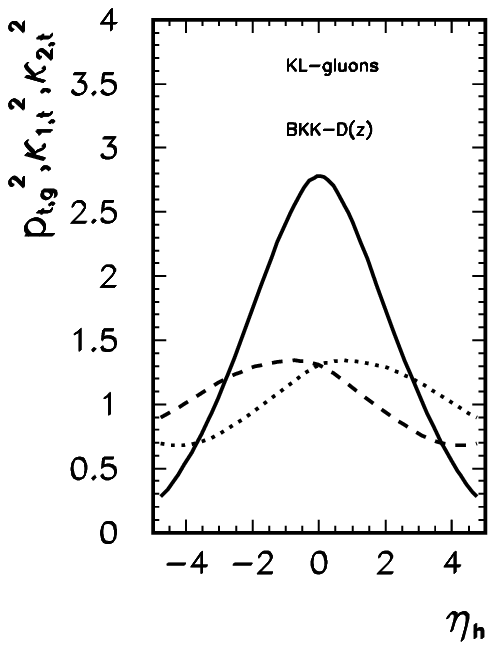}
\caption{\it Average values of
$\left< p_t^2 \right>$ (solid),
 $\left< \kappa_1^2 \right>$ (dashed) and
 $\left< \kappa_2^2 \right>$ (dotted)
as a function of pion pseudorapidity
\label{fig:ave_pt_eta}
}
\end{center}
\end{figure}

What are typical transverse momenta of gluons
involved in the calculations is shown in Fig.\ref{fig:ave_pt_eta}.
In this calculation we have used the KL unintegrated gluon
distribution and the BKK $g \rightarrow \pi$
fragmentation functions \cite{BKK95}.
We observe a maximum of the transverse momentum squared
of the produced gluon at $\eta_h \approx$ 0.
In our implementation of fragmentation ($p_{t,h}^2 \ll p_{t,g}^2$)
one tests relatively large $p_{t,g}^2$.
While at midrapidities $ \left< p_t^2 \right> >
\left< \kappa_1^2 \right>, \left< \kappa_2^2 \right> $,
when going to the fragmentation regions the relation reverses.
In the whole range of pseudorapidity one tests on average
$ \kappa_1^2, \kappa_2^2 \sim$ 1 GeV$^2$. One should remember,
however, that at the same time $\left<x_1\right>$ and
$\left<x_2\right>$ change dramatically when going from midrapidities
to the fragmentation region.

\begin{figure}[htb] 

\begin{center}
\includegraphics[width=8cm]{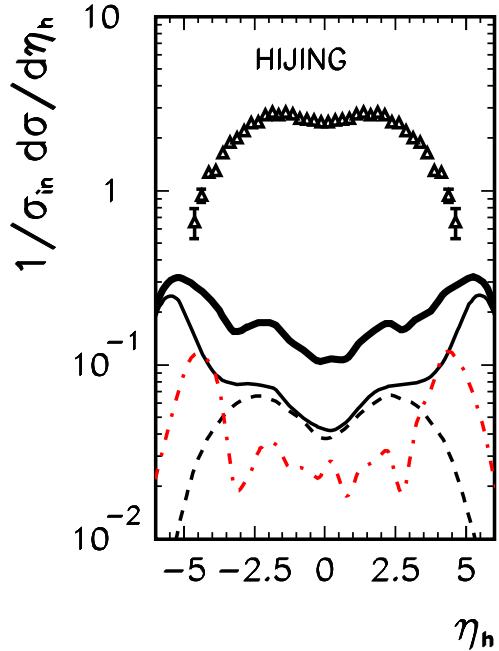}
\caption{\it
The pseudorapidity distribution of protons (solid), antiprotons
(dashed) and the difference of the spectra of $\pi^+$ and $\pi^-$
(dash-dotted) in the proton-proton collision at W = 200 GeV
obtained with the code HIJING \cite{HIJING}.
The thick solid line corresponds to the sum of these three
contributions.
The experimental data of the UA5 collaboration are taken
from \cite{UA5_exp}.
\label{fig:HIJING}
}
\end{center}
\end{figure}

In contrast to Ref.\cite{KL01}, where the whole pseudorapidity
distribution, including fragmentation regions, has been well
described in an approach similar to the one presented here,
in the present paper pions produced from the fragmentation
of gluons in the $gg \rightarrow g$ mechanism populate only
midrapidity region,
leaving room for other mechanisms in the fragmentation regions.
These mechanisms involve quark/antiquark degrees of freedom
or leading protons among others.
In Fig.\ref{fig:HIJING} I show the pseudorapidity spectra
of protons, antiprotons and the difference
$d \sigma/d \eta_{\pi^+} - d \sigma/d \eta_{\pi^-}$ obtained
with the code HIJING \cite{HIJING} (see also \cite{HIJING_had}).
The difference of the proton-antiproton spectra gives an idea
of leading particle contribution. Both protons from
deeply inelastic events as well as protons from diffraction dissociation
(single diffraction) have been included. The difference of
the positively and negatively charged pions gives the lower
limit on the $\pi^+ - \pi^-$ asymmetric mechanisms not taken
into account in the Kharzeev-Levin approach. The sum of the
three contributions (thick solid) gives then lower limit
on the missing contributions.
It is of the similar size
as the missing contributions in Fig.\ref{fig:eta_ff}
and Fig.\ref{fig:eta_glue}. This strongly suggests
that the agreement of the result of the $gg \rightarrow g$
approach with the PHOBOS distributions \cite{PHOBOS} in
Ref.\cite{KL01} in the true fragmentation region is rather due to
approximations made in \cite{KL01} than due to correctness
of the reaction mechanism. In principle, this can be verified
experimentally at RHIC by measuring the $\pi^+ / \pi^-$ ratio
in proton-proton scattering as a function of (pseudo)rapidity
in possibly broad range.
It seems that the BRAHMS experiment, for instance, can do it
even with the existing apparatus. 
 

\begin{figure}[htb] 

\begin{center}
\includegraphics[width=8cm]{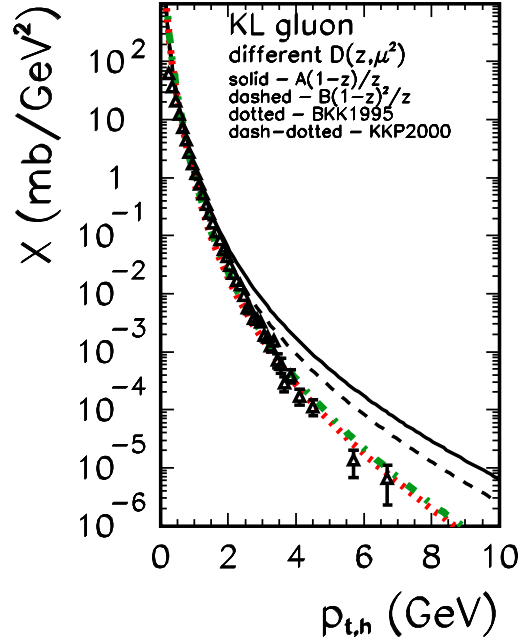}
\caption{\it
Transverse momentum distributions of charged pions at W = 200 GeV
for the KL gluon distribution and different fragmentation functions.
The experimental data of the UA1 collaboration are taken
from \cite{UA1_exp}.
\label{fig:pt_ff}
}
\end{center}
\end{figure}

The transverse momentum distribution of charged hadrons is shown
in Fig.\ref{fig:pt_ff} together with experimental data of
the UA1 collaboration at CERN from Ref.\cite{UA1_exp}.
In this calculation the KL gluon distribution has been used.
It is not completely clear to me how the
experimental data in \cite{UA1_exp} should be interpreted.
\footnote{The notion of the invariant cross section in \cite{UA1_exp}
is contradictory to the lack of particle identification there.}
I assume that the experimental data should be interpreted as:
\begin{equation}
X = \int \frac{d \sigma}{d \eta_h d^2 p_t} d \eta_h
\; / \; \int d \eta_h
\; .
\label{UA1_interpretation}
\end{equation}
We have taken $\eta_h \in$ (-2.5,2.5).
The simple hadronization functions, called model I and II above,
correctly fit low $p_{t,h}$ data and fail in the large $p_{t,h}$
region. This is due to lack of QCD evolution \cite{FF_evolution}.
The results obtained with fragmentation functions from
\cite{BKK95,KKP00} which include DGLAP evolution,
extremely well describe the large $p_{t,h}$ data.
Having in mind the ambiguity of the experimental data interpretation,
the KL gluon distribution does a fairly good job.

\begin{figure}[htb] 

\begin{center}
\includegraphics[width=8cm]{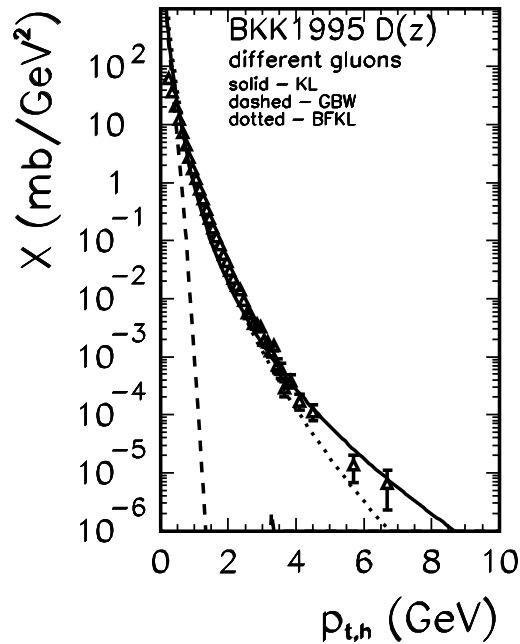}
\caption{\it
Transverse momentum distributions of charged pions at W = 200 GeV
for BKK1995 fragmentation function and different models of
unintegrated gluon distributions.
The experimental data of the UA1 collaboration are taken
from \cite{UA1_exp}.
\label{fig:pt_glue}
}
\end{center}
\end{figure}

In Fig.\ref{fig:pt_glue} I compare the theoretical transverse
momentum distributions of charged pions obtained with
different gluon distributions with the UA1 collaboration data
\cite{UA1_exp}. The best agreement is obtained with the
Karzeev-Levin gluon distribution. The distribution with
the GBW model is much too steep in comparison to experimental
data. This is probably due to neglecting QCD evolution.


\section{Conclusions}

I have calculated the inclusive distributions of gluons and
associated charged pions in the nucleon-nucleon collisions
through the $g g \rightarrow g$ mechanism
in the $k_t$-factorization approach. The results for several
unintegrated gluon distributions proposed recently in the literature
have been compared. The results, especially transverse
momentum distributions, obtained with different models of
unintegrated gluon distributions differ considerably.

A special attention has been devoted to the gluon distribution
proposed recently by Kharzeev and Levin to describe charged
particle production in relativistic heavy-ion collisions.
In the first step I have tested the gluon distribution
in electron deep-inelastic scattering at small Bjorken $x$.
A rather good description of the HERA data can be obtained
by adjusting a normalization constant.
In the next step so-fixed gluon distribution has been used
to calculate (pseudo)rapidity and transverse momentum distribution
of gluonic jets and charged particles.

Huge differences in both rapidity and transverse
momentum distributions of gluons and pions for different models
of unintegrated gluon distributions have been found.

Some approximations used recently in the literature
have been discussed. Contrary to a recent claim in Ref.\cite{KL01},
we have found that the gluonic mechanism discussed does not describe
the inclusive spectra of charged particles in the fragmentation
region, i.e. in the region of large (pseudo)rapidities for
any unintegrated gluon distribution from the literature.
Clearly the gluonic mechanism is not the only one and other
mechanisms (see e.g.\cite{W00,EH02} ) neglected in \cite{KL01}
must be added.
Some of them have been estimated with the help of the HIJING code,
giving a right order of magnitude for the missing strength.

Since the mechanism considered is not complete, it is not possible
at present to precisely verify different models of unintegrated gluon
distributions. The existing gluon distributions lead
to the contributions which almost exhaust the strength
at midrapidities and leave room for other mechanisms
in the fragmentation regions. It seems that a measurement of
transverse momentum distributions of particles at RHIC should
be helpful to test better different unintegrated gluon distributions.
A good identification of particles is required to verify
the other mechanisms.

In contrast to standard integrated gluon distributions,
the extraction of unintegrated gluon distribution
from experimental data seems a rather difficult task.
At present, one can rather test different unintegrated
gluon distributions based on different models existing
in the literature. In the present analysis I have discussed
whether the production of particles can provide some information
on unintegrated gluon distributions in the nucleon.
There are many other reactions where this is possible,
to mention here only heavy quark or jet production
in $e p$ and $p p$ collisions.
Going to more exclusive measurements seems indispensable.
An example is a careful study of jet correlation
in photon-proton \cite{SNSS01} and nucleon-nucleon \cite{LO00}
collisions. In my opinion, we are at the beginning of
the long way to extract gluon or more generally parton unintegrated
distributions.

\vskip 1cm

{\bf Acknowledgements}
I am indebted to Jan Kwieci\'nski for several
discussions on different subjects concerning high-energy physics,
for his willingness to share his knowledge, for his optimism
and friendly attitude.
The discussion with Andrzej Budzanowski is kindly acknowledged.
The pseudorapidity distributions of particles from the code HIJING
have been obtained from Piotr Paw{\l}owski.
I thank Leszek Motyka for providing me with some FORTRAN routines.



\end{document}